\documentclass[letter]{aa}  

\usepackage{graphicx}
\bibpunct{(}{)}{;}{a}{}{,}
\usepackage{txfonts}
\usepackage{amssymb}
\usepackage{multirow}
\usepackage{float}
\usepackage{gensymb}
\usepackage{color}
\usepackage{rotating}
\usepackage{hyperref}
\hypersetup{backref=true, pagebackref=true, hyperindex=true, breaklinks=true,colorlinks=true,urlcolor=blue, linkcolor=blue, citecolor=blue,pagecolor=red, bookmarks=true, bookmarksopen=true}

\begin{document}

   \title{Observing the linked depletion of dust and CO gas\\ at 0.1-10\,au in disks of intermediate-mass stars}


   \author{A.\,Banzatti \inst{\ref{Tucson}}
   \and A.\,Garufi\inst{\ref{Cantoblanco}}
   \and M.\,Kama\inst{\ref{Cambridge}}
   \and M. Benisty \inst{\ref{Myriam1},\ref{Myriam2}}
   \and S.\,Brittain\inst{\ref{Clemson}}
   \and K. M. Pontoppidan \inst{\ref{STScI}}
   \and J.\,Rayner\inst{\ref{Hawaii}}
          }

\institute{Lunar and Planetary Laboratory, The University of Arizona, Tucson, AZ 85721, USA \label{Tucson}
\email{banzatti@lpl.arizona.edu}
 \and Universidad Auton\'{o}noma de Madrid, Dpto. F\'{i}sica Te\'{o}rica, Facultad de Ciencias, Campus de Cantoblanco, E-28049 Madrid, Spain\label{Cantoblanco}
 \and Institute of Astronomy, Madingley Rd, Cambridge, CB3 0HA, UK \label{Cambridge}  
 \and Unidad Mixta Internacional Franco-Chilena de Astronom\'{i}a, CNRS/INSU UMI 3386 and Departamento de Astronom\'{i}a, Universidad de Chile, Casilla 36-D, Santiago, Chile.  \label{Myriam1}
 \and Univ. Grenoble Alpes, CNRS, IPAG, 38000 Grenoble, France  \label{Myriam2}
 \and Department of Physics \& Astronomy, 118 Kinard Laboratory, Clemson University, Clemson, SC, USA \label{Clemson}     
 \and Space Telescope Science Institute, 3700 San Martin Drive, Baltimore, MD 21218, USA \label{STScI} 
 \and University of Hawaii, 2680 Woodlawn Drive, Honolulu, HI, USA 96822-0345 \label{Hawaii}     
             }

   \date{Received - / Accepted November 26, 2017}

 \abstract 
{
We report on the discovery of correlations between dust and CO gas tracers of the 0.1--10 au region in planet-forming disks around young intermediate-mass stars. The abundance of refractory elements on stellar photospheres decreases as the location of hot CO gas emission recedes to larger disk radii, and as the near-infrared excess emission from hot dust in the inner disk decreases. The linked behavior between these observables demonstrates that the recession of infrared CO emission to larger disk radii traces an inner disk region where dust is being depleted. We also find that Herbig disk cavities have either low ($\sim$\,5--10\,\%) or high ($\sim$\,20--35\,\%) near-infrared excess, a dichotomy that has not been captured by the classic definition of ``pre-transitional'' disks. 
} 
 

\authorrunning{Banzatti et al.\,2017}

\titlerunning{Observing the linked depletion of dust and CO gas}

   \maketitle


\section{Introduction} \label{sec: intro}
The vast majority of exoplanets discovered so far lies at less than 3 au from the central star, with super-Earths abundant well inside 1 au \citep[e.g.,][]{petigura13}. Planets of up to a few Earth masses could form in situ \citep[e.g.,][]{hansen12}, while giant planets most probably migrate inward from beyond a few au \citep[e.g.,][]{KleyNelson12}.
Whether exoplanets form or migrate where they are detected, the effect of the structure and evolution of protoplanetary disks at 0.1--10\,au is expected by all models to be fundamental in shaping planetary systems. 

First measurements on the evolution of inner disks came from spatially unresolved observations of spectral energy distributions (SEDs), which in some disks show a lower infrared (IR) flux that
was attributed to a deficit of hot inner material \citep{strom1989}. This deficit has been interpreted as due to inner holes (``transitional'' disks) or gaps (``pre-transitional'' disks), depending on the level of near-IR flux related to an inner dust belt inside the cavity \citep{espaillat2007}. Modern imaging techniques, probing disk radii of $\gtrsim 5$\,au, have confirmed the existence of $> 10$\,au wide cavities in several disks, depleted of dust particles of up to centimeter sizes \citep[e.g.,][]{andrews2011}. A leading theory for the origin of these cavities is disk clearing by planets, exo-Jupiters, but possibly also super-Earths \citep[e.g.,][]{pinilla12,fung17}. Inner disks can be dispersed by winds as well, although observations cannot yet be fully reconciled in any photoevaporative or planet-disk interaction models \citep{owen16}. In disks around young intermediate-mass stars (Herbig Ae/Be stars), the far-IR excess of the SED, tracing colder material at larger radii, has been adopted by \citet{meeus01} to classify disks into Group I (GI, with high excess) and Group II (GII, with moderate excess). The current understanding of this empirical classification is that it reflects a different disk structure, with GI having a large disk cavity that allows the irradiation of the disk at $\gtrsim 10$ au, and GII having no or at most small inner cavities \citep{maask13,menu15,honda15,garufi17}.    
 
While near-IR and millimeter imaging now reveal increasing detail in global disk structures \citep[e.g.,][]{alma15,benisty2015}, information on the inner disk structure at $\lesssim 5$ au can only be obtained from optical and near-IR spectroscopy and near-IR interferometry. Recently, independent studies have found new evidence in dust and gas tracers pointing in the direction of depletion processes in these inner disk regions \citep{kama15,bp15}. By combining these independent datasets (Sect.\,\ref{sec: data}), we report in this work on the discovery of a linked behavior between observables of CO gas and dust in inner disks (Sect.\,\ref{sec: correlations}). This behavior demonstrates a strong link between molecular gas and dust, providing an important framework for better understanding the evolving structure of planet-forming regions at $\lesssim 10$ au (Sect.\,\ref{sec:disc}).

\begin{figure*}
\centering
\includegraphics[width=1\textwidth]{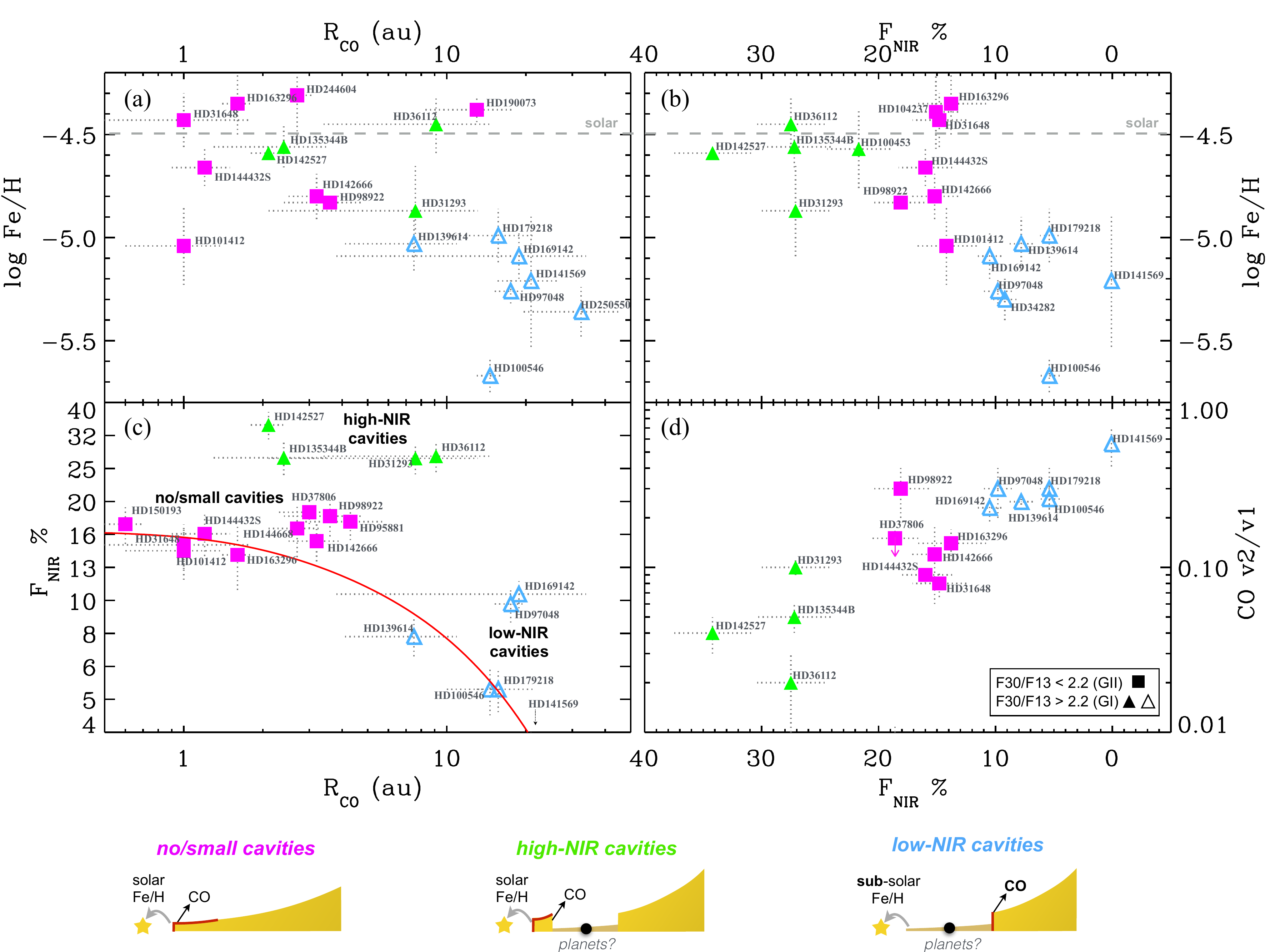} 
\caption{Linked behavior between the datasets combined in this work. \textbf{(a)}: Fe/H vs R$_{\rm co}$. \textbf{(b)}: Fe/H vs $\rm F_{NIR}$. \textbf{(c)}: $\rm F_{NIR}$ vs R$_{\rm co}$. \textbf{(d)}: $v2/v1$ vs $\rm F_{NIR}$. The red curve shows a parametric model of the decrease of $\rm F_{NIR}$ with increasing size of an inner cavity (see text for details). The three disk categories identified in the multi-dimensional parameter space are illustrated to the bottom. Dust is shown in yellow, and we mark the approximate location of infrared CO emission. Dust depletion is shown as a thinner yellow layer of residual dust, and dust layers that dominate the observed $\rm F_{NIR}$ are marked in red. GII disks are shown as magenta squares, high-NIR GI disks as green triangles, low-NIR GI disks as cyan (empty) triangles.}
\label{fig: correlations}
\end{figure*}


\begin{figure*}[ht]
\centering
\includegraphics[width=1\textwidth]{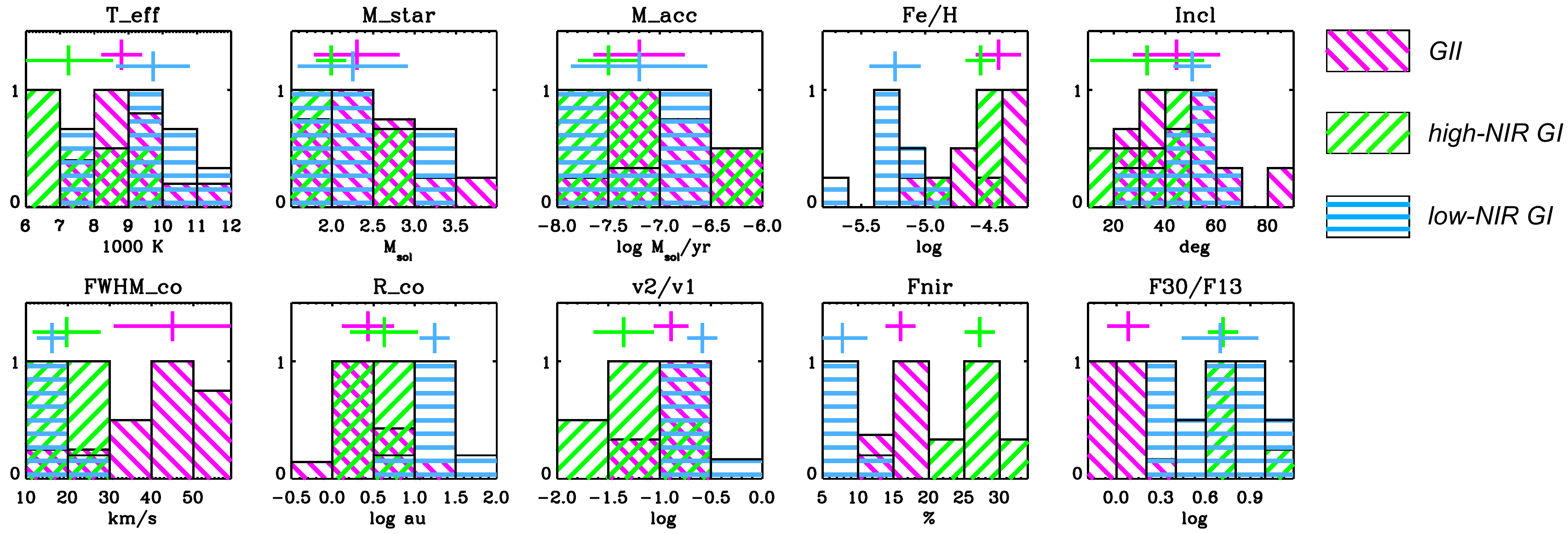} 
\caption{Histograms of normalized distributions of the sample parameters included in Table \ref{tab: sample}. Median values and median absolute deviations for each category are plotted at the top of each histogram. GII disks are shown in magenta, high-NIR GI disks in green, and low-NIR GI disks in cyan. }
\label{fig: histos}
\end{figure*}

\section{Sample and Measurements} \label{sec: data}

The three independent datasets combined in this analysis are the orbital radius and excitation of rovibrational CO emission (Sect. \ref{sec: CO}), the iron abundance Fe/H as measured on stellar photospheres (Sect. \ref{sec: Fe}), and the near-IR excess over the stellar flux in the SED (Sect. \ref{sec: NIR}).
The sample includes the majority of well-known Herbig Ae/Be stars within 200 pc, together with some at larger distances: 26 late-B, A, and F stars with temperatures $T_{\rm eff}$ between 6,500 K and 11,000 K, masses between 1.5 M$_{\odot}$ and 4 M$_{\odot}$, and ages between 1 and 10 Myr. The sample is composed evenly by GI and GII disks (13 each), and we adopt the classification criterion from \citet{garufi17}, where the flux ratio at 30 and 13 $\mu$\,m is used to separate GI disks (with $F30/F13 > 2.2$) from GII disks
(with $F30/F13 < 2.2$, where 2.2 is the ratio for a flat SED). All three datasets are available for 16 of these objects, while only two are available for the other 10 (Table \ref{tab: sample}).

\subsection{Rovibrational CO emission} \label{sec: CO}
Spectrally resolved emission line profiles provide information on gas kinematics in protoplanetary disks. Rovibrational CO emission at 4.7--4.8\,$\mu$m was used to estimate a characteristic orbital radius of hot/warm ($\sim$300--1500 K) CO gas in Keplerian rotation in a disk, as $\rm{R}_{co} = (2\, \rm sin \,\textit{i}\, /\, FWHM_{\rm co})^2 \, G\,M_{\star}$, where M$_{\star}$ is the stellar mass and $i$ the disk inclination. As a characteristic gas velocity, we took the CO line velocity at the half width at half maximum (FWHM$_{\rm co}/2$) as in \citet{bp15}. For the disk inclinations, we adopted values from the near-IR interferometric survey by \citet{lazareff17}, which probe inner disks at $< 10$\,au. The 1$\sigma$ errors on R$_{\rm co}$ were propagated from the uncertainties on stellar masses, CO line widths, and disk inclinations, and they are dominated by the disk inclination uncertainties.
Most CO spectra have been published previously, except for four disks that were newly observed with IRTF-ISHELL (see Appendix \ref{sec:ishellspec}).
Another measured property of CO emission is the flux ratio between rovibrational lines from the second and first vibrational levels. The ratio $v2/v1$ is a sensitive tracer of the type of CO excitation \citep[e.g.,][]{britt03,britt07,thi13}: UV-pumping populates high vibrational states first (producing higher $v2/v1$), while IR-pumping and collisional excitation populate low states first (producing lower $v2/v1$).

\subsection{Refractory abundance on stellar photosphere} \label{sec: Fe}
Modeling of high-resolution optical spectra enables measuring elemental abundances on stellar photospheres. We used the stellar Fe/H compilation from \citet{kama15}, including mostly values from \citet{folsom12}. The very shallow surface convection zone in stars with $\rm T_{eff}\gtrsim7500$\,K inhibits mixing with the bulk of the stellar envelope and keeps accreted material visible on the photosphere for $\sim1\,$Myr. The Fe/H measurements in Herbig~Ae/Be stars have recently been found to correlate with the presence or absence of dust cavities detected by millimeter interferometry imaging, suggesting that the stellar photospheres keep an imprint of the dust/gas ratio of their inner disks through the accreted material \citep{kama15}. Two stars in the sample have $\rm T_{eff}$ lower than 7500 K, HD142527 and HD135344B; these stars are likely mixing the accreted material more efficiently than the rest of the sample.

\subsection{Near-infrared excess} \label{sec: NIR}
A traditional probe of hot dust in inner disks is the near-IR excess \citep[e.g.,][]{dominik03}. We estimated the fractional $\rm F_{NIR}$ = F(NIR)/F$_\star$ for our entire sample to ensure an homogeneous procedure, and found values consistent with those estimated in previous work. We collected the BVRJHK photometry, along with the WISE fluxes at 3.6 $\mu$m and 4.5 $\mu$m and dereddened by means of the extinction $A_{\rm V}$ available from the literature \citep[$A_{\rm V} < 0.5$\,mag for most of this sample;][]{meeus12}. A PHOENIX model of the stellar photosphere \citep{Hauschildt1999} with the literature stellar temperature and metallicity and a
surface gravity log$(g)=-4.0$ was employed for each source and scaled to the dereddened V magnitude for each source. The near-IR excess $\rm F_{NIR}$ was measured by integrating the observed flux exceeding the stellar flux between 1.2 $\mu$m and 4.5 $\mu$m. These values were then divided by the total stellar flux F$_\star$ from the model. The 1$\sigma$ errors on $\rm F_{NIR}$, propagated from the uncertainties on $\rm T_{eff}$ and an assumed uncertainty of 0.2 mag in $A_{\rm V}$, are typically 13\% (median value), and always $\lesssim 20\%$ (Table \ref{tab: sample}).

\section{Linked behavior between the datasets} \label{sec: correlations}
The datasets combined in this work show a linked behavior in the multi-dimensional parameter space illustrated in the four panels of Figure \ref{fig: correlations}.
The correlation between Fe/H and R$_{\rm co}$ demonstrates a link between two observables that could in principle be independent of each other: iron is depleted from the stellar photospheres as R$_{\rm co}$ recedes to larger radii in their inner disks. 
GII disks have, on average, smaller R$_{\rm co}$ and higher Fe/H, while GI disks have the opposite, larger R$_{\rm co}$ and lower Fe/H. A group of GI disks overlaps with GII disks at intermediate values of R$_{\rm co}$. 

$\rm F_{NIR}$ values are found to lie between 5\% and 34\%, with only one disk showing a value as low as 0.08\% (HD141569, see Sect.\,\ref{sec: environment}). All GII disks show $\rm F_{NIR}$ in the narrow range between 14\% and 19\%. GI disks instead span
a wider range of $\rm F_{NIR}$, some with lower values than the GII (5--11\%) and some with higher values \citep[22--34\%, see also][]{garufi17}. Overall, disks with larger R$_{\rm co}$ have lower $\rm F_{NIR}$ and Fe/H, although without a simple monotonic relation between R$_{\rm co}$ and $\rm F_{NIR}$. In addition, there is a strong linear anticorrelation between $\rm F_{NIR}$ and the CO vibrational ratio $v2/v1$ (Fig. \ref{fig: correlations}d).
 
Three categories of disks can be identified as based on inner disk observables (Figure \ref{fig: correlations} and Table \ref{tab: categories}). Beyond nomenclature or classification intents, we can understeand the evolving properties of inner disks more
comprehensively by identifying which observable properties separate out in this multi-dimensional space. Interestingly, the GI disks in the sample are split into two very distinct parts of the parameter space, and we refer to them as ``high-NIR'' and ``low-NIR'' GI disks to identify the different behavior of inner disk observables. This dichotomy in $\rm F_{NIR}$ values of GI disks is strongly segregated, with the Kolmogorov-Smirnov (KS) two-sided test highly rejecting the hypothesis that they may be drawn from a same parent distribution (probability of < 1\%). The three categories are instead not separated in terms of stellar temperature, mass, luminosity, or age, nor in mass accretion rates (see Figure \ref{fig: histos}), and no correlation is found between these parameters and R$_{\rm co}$, $\rm F_{NIR}$, or Fe/H, suggesting that the measured behavior of inner disk observables cannot be attributed to these parameters.

\begin{table}
      \caption[]{Median values for the three disk categories  as in Fig.\,\ref{fig: histos}.}
         \label{tab: categories}
     $$ 
         \begin{tabular}{cccccc}
\hline
            \hline
            \noalign{\smallskip}
            log(Fe/H) & $\rm R_{co}$ & $\rm F_{NIR}$ & $v2/v1$ & $F30/F13$ & Cavity? \\
            \hline
            \noalign{\smallskip}
            
            $-4.4$ & $3$ au & $16\%$ & $0.12$ & $1.2$ & no/small \\ 
            \noalign{\smallskip}
            
            $-4.6$ & $5$ au & $27\%$ & $0.05$ & $5$ & ``high-NIR'' \\ 
            \noalign{\smallskip}
            
            $-5.2$ & $18$ au & $8\%$ & $0.27$ & $5$ & ``low-NIR'' \\ 
             \noalign{\smallskip}
            \hline
            
\end{tabular}
$$
\end{table}

The linked behavior between inner disk tracers of CO and dust may be produced by a scenario where as dust is depleted (as shown by the decrease in $\rm F_{NIR}$ and Fe/H), CO gas is depleted as well (CO emission at high velocity decreases, FWHM$_{\rm co}$ shrinks, and $\rm R_{co}$ moves to larger disk radii). As a simple test of whether a linked depletion of CO gas and dust from the inner disk may globally reproduce the observed trends, we adopted a parametric description of an inner disk structure in hydrostatic equilibrium, as based on work by \citet{kama09} and summarized in Appendix \ref{sec:model}. The disk temperature decreases with disk radius as $r^{-0.5}$, and we explored how $\rm F_{NIR}$ decreases as the radius of the innermost dust (R$_{\rm dust}$) was sequentially increased to mimic the formation of an inner hole. 

A decrease of $\rm F_{NIR}$ with increasing hole size (red curve in Fig.\,\ref{fig: correlations}c) is naturally produced by a decreasing temperature of the emitting dust, as hotter dust at smaller radii is removed. In this scenario, R$_{\rm co}$ could trace the size of an inner dust-depleted disk region, as suggested from previous observations \citep{britt03,britt07}. When Keplerian line profiles are modeled assuming a power-law radial brightness \citep[e.g.,][]{salyk11}, R$_{\rm co}$ as defined here can be several times larger than R$_{\rm dust}$ (taken as the innermost radius where CO gas can also exist). 
The model curve in Fig.\,\ref{fig: correlations}c shows R$_{\rm dust}$ multiplied by 7, by assuming the median scaling factor between R$_{\rm co}$ values and the R$_{\rm dust}$ values from \citet{lazareff17} for this sample. By assuming devoid inner holes, this simple model provides only a trend of the lower boundary to the data points in the figure. Reproducing the high near-IR excess of some Herbig disks has been a long-standing modeling challenge that is still unsolved \citep[see, e.g.,][]{dullemond01,vinko06}. New modeling efforts could be devoted to studying how $\rm F_{NIR}$ can be increased toward values measured in the high-NIR GI disks (25--35\%) by keeping an inner residual dust component while inner cavities are forming at larger disk radii. Such a structure is supported by recent observations of these disks, as discussed below.


\section{Discussion} \label{sec:disc}

\subsection{Dust and gas depletion in the inner disk regions} \label{sec: gas_dust_depl}
The behavior of inner disk observables presented in this work can be affected by processes that involve different modifications to the radial distributions of dust and gas. Beyond the scope of this work, quantification of these processes requires explorations with sophisticated thermo-chemical models of disks, which have only recently started to be applied to disks with inner cavities \citep[e.g.,][]{simon13}. We advocate in this work a strong empirical evidence for a link in the evolution of dust and molecular gas in planet-forming regions, extending to inner disk radii smaller than can be spatially resolved by ALMA.

Sub-solar Fe/H (and other refractories) on radiative stellar photospheres have long been suggested to be linked to dust-depleted material accreted onto the star \citep[e.g.,][]{venn90}, but it was unclear why only some Herbig Ae/Be stars exhibit a depletion of refractories \citep{folsom12}. \citet{kama15} showed that sub-solar refractory abundances in Herbig stars correlate with the presence of a large ($>$ 10 au) dust cavity in their disks, pointing out that the depletion of dust grains in the inner regions of transitional disks would naturally explain a lack of refractories in the accreting material. The correlation between Fe/H and $\rm F_{NIR}$ shown here demonstrates that a link exists between refractory abundances on stellar photospheres and the properties of dusty inner disks at $<10$ au. To decrease both Fe/H and $\rm F_{NIR}$, dust must be depleted from both the accretion flow \textit{\textup{and}} from the inner disk. Potential processes include 1) dust grain growth/inclusion into solids larger than $\gg 1$ mm (pebbles and planetesimals) that decouple from the accreted gas and emit less efficiently in the infrared, 2) physical decoupling of inner and outer disk, where the resupply of inner disk dust by inward drift from the outer disk is inhibited \citep[see, e.g., discussions in][]{andrews2011,kama15}. 

The survival of CO gas in a dust-depleted cavity of Herbig disks was specifically investigated by \citet{simon13}. The study showed that as long as inner disks are still gas rich (N(H$_2$) $\approx10^{27}$cm$^{-2}$ at <1\,au), the radial distribution of CO gas does not depend on the presence of dust, as the density of CO molecules is well above the column needed to self-shield from UV photodissociation \citep{visser09}. When instead the \textit{\textup{total}} gas content in the inner disk is decreased by at least four orders of magnitude, reaching column densities where UV photodissociation of CO becomes relevant, the survival of CO is closely linked to the presence of shielding dust. Specifically, in a dust- \textit{\textup{and}} gas-depleted cavity, the CO column density decreases by orders of magnitude by UV photodissociation inside the cavity when a residual inner dust
belt is removed. In this situation, R$_{\rm co}$ will shift to larger radii as the dust is depleted from the innermost disk radii. In the framework of this modeling, the linked behavior between dust and CO gas observables suggests that low-NIR GI disks with large R$_{\rm co}$ may be in a \textit{\textup{gas-depleted}} regime. Gas depletion factors of 10$^2$--10$^4$ have been supported by recent analyses of millimeter CO emission in some of these disks, as imaged by ALMA \citep{vdmarel16}.

\begin{figure}
\centering
\includegraphics[width=0.5\textwidth]{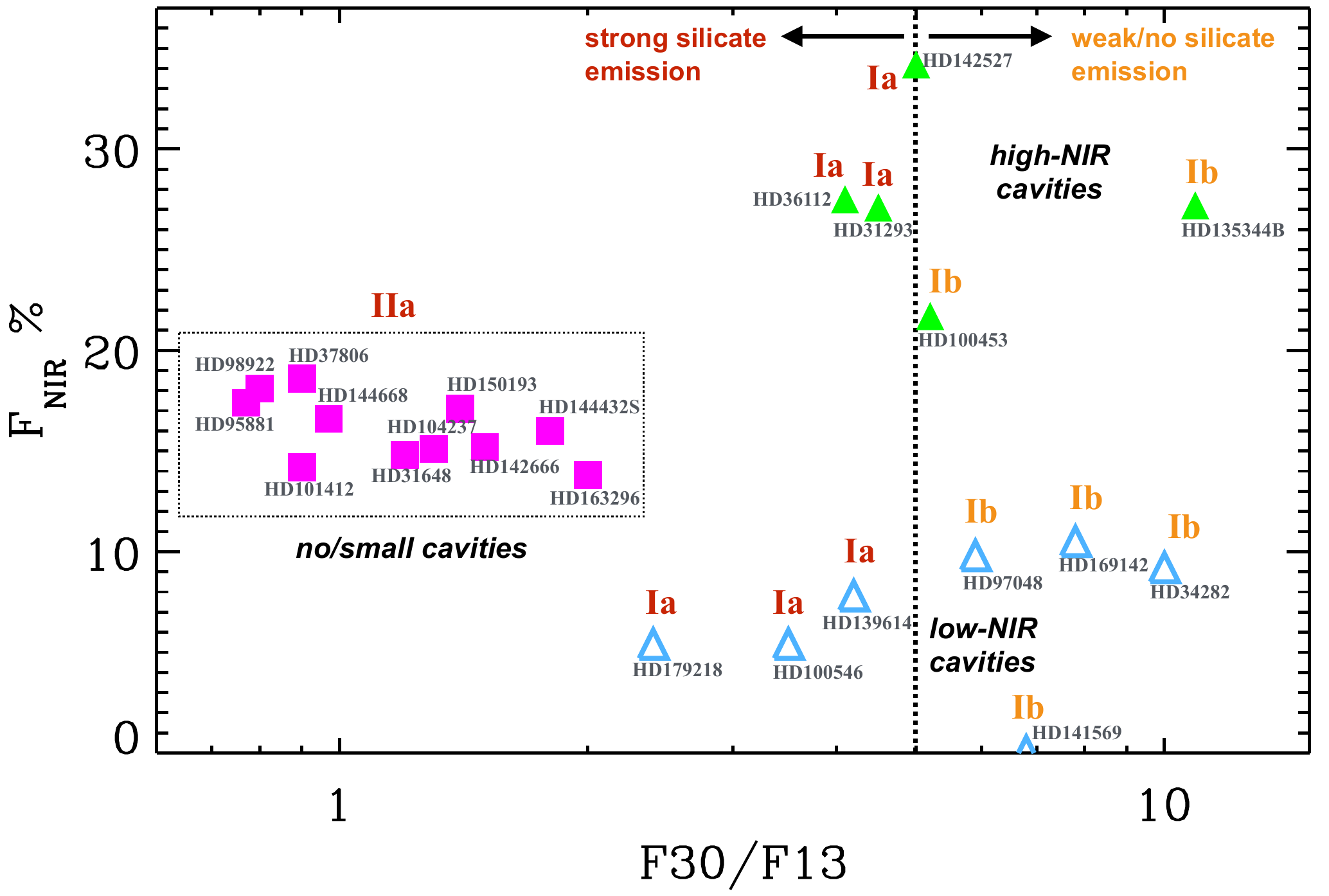} 
\caption{$\rm F_{NIR}$ and $F30/F13$ for this sample of Herbig disks, highlighting the three classes as in Figure \ref{fig: correlations}. Following \citet{meeus01}, we mark disks where the 10\,$\mu$m silicate emission has been detected (``IIa'' and ``Ia'') or not (``Ib''); we keep the color-coding for the Ia/Ib labels from \citet{maask13} to facilitate comparison to that work. \citet{maask13} showed that silicate emission is tightly related to the $F30/F13$ ratio; instead, $\rm F_{NIR}$ is unrelated (Sect. \ref{sec: environment}).}
\label{fig: Fnir_F30}
\end{figure}

\subsection{Dichotomy of inner disk cavities} \label{sec: environment}
The net separation between high-NIR and low-NIR GI disks suggests that large inner disk cavities have two possible structures. Their properties are at the opposite extremes of those of GII disks with no/small cavities (Figures \ref{fig: correlations} and \ref{fig: Fnir_F30}). This dichotomy once again suggests a strong link between dust and gas, consistent with the inner disk structure discussed above in thermo-chemical modeling work: a residual inner dust component (high $\rm F_{NIR}$) may allow for UV-shielded and vibrationally colder CO gas to survive in a dust-free cavity at larger radii ($v2/v1 < 0.1$ in high-NIR disks), while its removal would allow for an efficient UV photodissociation of CO gas inside the dust cavity and for UV pumping at the cavity wall ($v2/v1 \sim 0.3$ in low-NIR disks).

We note that $\rm F_{NIR}$ does not show a monotonic relation with $F30/F13$ (Figure \ref{fig: Fnir_F30}) and that the 10\,$\mu$m silicate emission is found in disks with cavities regardless of the measured level of $\rm F_{NIR}$. While $F30/F13$ traces the depletion of intermediate disk regions where the warm ($\sim$\,200--400\,K) silicate emission is produced \citep{maask13}, $\rm F_{NIR}$ in the high-NIR disks must be produced by a hotter disk region closer to the star.
We also note that no Herbig disk in this sample has $\rm F_{NIR} \sim 0,$ except for HD141569, which has been proposed to be globally dispersing its gas toward the debris disk phase \citep[e.g.,][]{White2016}. This suggests that inner cavities in Herbig disks are depleted but never completely devoid of material, consistent with residual dust detected at the sublimation radius by NIR interferometry \citep{lazareff17} and with significant accretion rates measured even in low-NIR disks. The moderate/high accretion rates measured in some stars with large inner disk cavities currently provide one of the main open problems in understanding the structure and origin of disk cavities \citep[e.g.,][]{owen16,EP17}; interestingly, CO gas depletion in the disk has recently been proposed as a potential solution \citep{ercolano17}.

The nature of a hot inner dust component in high-NIR cavities remains to be determined. Studies have investigated how $\rm F_{NIR}$ can be increased by i) increasing the inner rim scale-height by thermal or magnetic processes \citep{dullemond01,flock17}, ii) a dusty wind launched close to the dust sublimation radius \citep[e.g., HD31293 as modeled by][]{bans12}, iii) misaligned inner disks with warps induced by a gap-opening companion \citep{owen17}. All these scenarios invoke vertically extended inner dusty structures that will have effects on CO gas excitation, and still need to be investigated by thermo-chemical models. Intriguingly, the presence of misaligned inner disks has been linked to the presence of shadows and spirals at larger disk radii \citep[e.g.,][]{montesinos2016, min17}, which have been imaged in \textit{\textup{all}} high-NIR GI disks \citep[][]{benisty2015,stolker16,avenh17,benisty17, tang2017}.  

\begin{figure}
\centering
\includegraphics[width=0.5\textwidth]{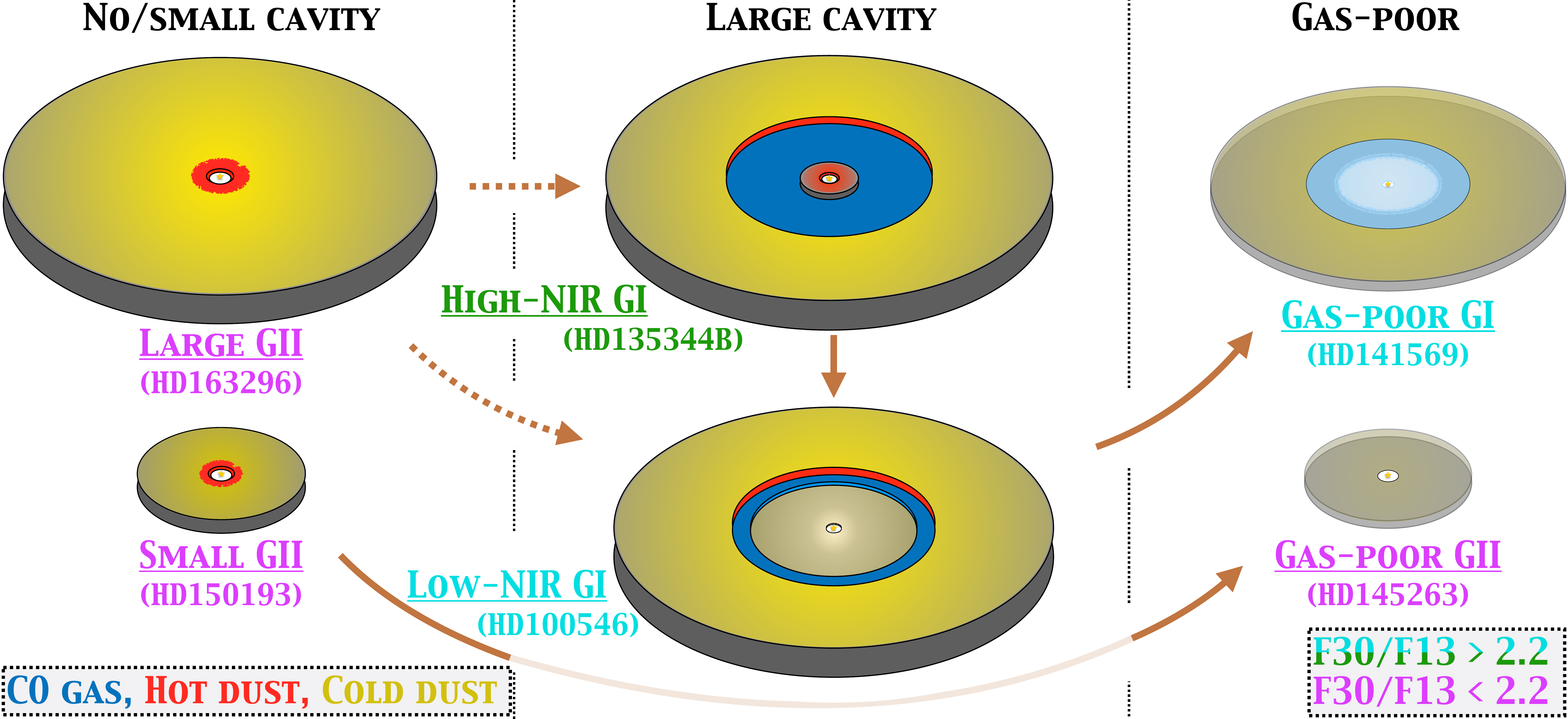} 
\caption{Evolutionary paths that appear possible by combining this work and the analysis of \citet{garufi17}. The dichotomy in inner disk cavities is shown at the center: high-NIR cavities have a residual inner dust component, possibly a belt/warp, that shields UV radiation and enables CO (vibrationally cold) to survive into the cavity; low-NIR cavities are instead dust- and gas-depleted, and CO (here vibrationally hot) is detected only close to the cavity wall.}
\label{fig: cartoon}
\end{figure}

We highlight that the dichotomy of inner disk cavities presented in this work has not been captured by the classic definition of a ``pre-transitional'' disk, as low-/high-NIR GI disks have been indistinguishably classified ``pre-transitional'' \citep{espaillat2014}. A combination of multiple tracers of dust and gas in inner and outer disk regions is needed to refine previous concepts of ``(pre-)transitional'' disks into a better understanding of their structure and evolutionary phase.
While some GII disks are already too small in radius and low in mass to ever show the properties of GI disks \citep[e.g., HD150193 and HD145263; see][]{garufi17}, it is possible that others are still overall large and massive enough to do so, if they eventually form a large inner cavity (e.g., the GII disk HD163296; Fig. \ref{fig: cartoon}). As argued in previous work, this suggests that at least two paths of disk evolution may be sampled by current observations. The observed dichotomy of GI cavities may instead suggest a potential ``on/off'' behavior of the hot inner dust component, where as soon as an inner dust belt/warp is removed, R$_{\rm co}$ abruptly recedes to $\gtrsim 10$\,au due to the destruction of residual CO gas by UV photodissociation in a dust- and gas-depleted cavity (Sect. \ref{sec: gas_dust_depl}).

\section{Conclusions} \label{sec: conclusions}
From the combination of three independent tracers of dust and CO gas in the inner disks of intermediate-mass stars, we conclude
that \begin{itemize}
\item the recession of NIR CO emission to larger disk radii traces dust depletion in inner disks at $\approx$\,0.1--10\,au, providing key measurements of disk evolution in inner regions beyond reach of direct-imaging techniques; based on recent thermo-chemical modeling \citep{simon13}, this behavior seems to imply that these disk cavities may also be gas depleted;

\item the multi-dimensional space of the several observables now available suggests that large cavities in Herbig disks form with either low ($\sim$\,5--10\,\%) or high ($\sim$\,20--35\,\%) NIR excess, a dichotomy that was not captured by the classic definition of ``pre-transitional'' disks; high-NIR GI disks seem to have residual inner dust belts/warps that have recently been inferred also from shadows and spirals at larger disk radii.
\end{itemize}


\begin{acknowledgements}
We thank A. Carmona for providing CO data of HD169142, and acknowledge helpful feedback and discussions with A. Bosman, G. Dipierro, U. Gorti, G. Mulders, P. Pinilla, and E. van Dishoeck. We also thank the anonymous referee for suggestions that helped improve the presentation of this work. M.B. acknowledges funding from ANR of France under contract number ANR-16-CE31-0013 (Planet Forming Disks). This work is partly based on observations obtained at the Infrared Telescope Facility, which is operated by the University of Hawaii under contract NNH14CK55B with the National Aeronautics and Space Administration. This research has made use of the VizieR catalogue, CDS, Strasbourg, France (A\&AS 143, 23).
\end{acknowledgements}

\begin{appendix}

\section{The sample} \label{Appendix_sample}
The sample properties are reported in Table \ref{tab: sample}.

\begin{table*}
      \caption[]{Sample properties, separated into the three disk categories discussed in the text. Median values and median absolute deviations (in parentheses) of all parameters are included below each disk category. Stellar temperatures and masses are from \citet{folsom12} and \citet{fair15}; accretion rates are from \citet{fair15}; Fe/H values are taken from the compilation in \citet{kama15}. Inner disk inclinations are adopted from \citet{lazareff17}, except for HD101412 \citep{fed08}, HD141569 \citep{White2016}, and HD135344B \citep{stolker17}. CO line widths and vibrational ratios are taken from the compilation in \citet{banz17} and from this work, except for HD169142, which is taken from Carmona et al. (in prep.). R$_{\rm{co}}$ and $\rm F_{NIR}$, as well as their uncertainties given in parentheses, are measured as explained in Section \ref{sec: data}. $F30/F13$ values are adopted from \citet{maask14}. }
         \label{tab: sample}
     $$ 
         \begin{tabular}{lcccccccccc}
            \hline
            \hline
            \noalign{\smallskip}
            Name & T$_{\rm eff}$ & M$_{\star}$ & M$_{acc}$ & log(Fe/H) & incl & FWHM$_{\rm co}$ & R$_{\rm{co}}$ & $v2/v1$ &  $\rm F_{NIR}$ &  $\rm F30/F13$ \\
             & (K) & (M$_{\odot}$) & (M$_{\odot}$/yr) & & (deg) & (km/s) & (au) & & (\%) & \\
            \hline
            \noalign{\smallskip}

HD31648 & 8800 & 2.1 & -6.9 & -4.43 & 39. & 55. & 1.0 (0.8) & 0.08 & 14.8 (2.3) & 1.2 \\
HD37806 & 11000 & 3.9 & -6.3 & -- & 41. & 45. & 3.0 (0.7) & <0.15 & 18.6 (2.0) & 0.9 \\
HD95881 & 9000 & 2.0 & -- & -- & 52. & 32. & 4.3 (1.4) & -- & 17.4 (2.1) & 0.8 \\
HD98922 & 10500 & 4.0 & -- & -4.83 & 20. & 22. & 3.6 (1.2) & 0.30 & 18.1 (2.4) & 0.8 \\
HD101412 & 8600 & 3.0 & -7.0 & -5.04 & 80. & 100. & 1.0 (0.4) & -- & 14.2 (2.6) & 0.9 \\
HD104237 & 8000 & 2.0 & -6.7 & -4.39 & -- & -- & -- & -- & 15.1 (1.9) & 1.3 \\
HD142666 & 7500 & 2.0 & -7.8 & -4.80 & 60. & 40. & 3.2 (0.7) & 0.12 & 15.2 (2.0) & 1.5 \\
HD144432S & 7400 & 2.0 & -7.4 & -4.66 & 25. & 32. & 1.2 (0.3) & 0.09 & 16.0 (2.3) & 1.8 \\
HD144668 & 8500 & 2.5 & -6.3 & -- & 52. & 45. & 2.7 (0.6) & -- & 16.6 (3.0) & 1.0 \\
HD150193 & 9000 & 1.9 & -7.5 & -- & 32. & 54. & 0.6 (0.1) & -- & 17.1 (2.2) & 1.4 \\
HD163296 & 9200 & 2.3 & -7.5 & -4.35 & 48. & 53. & 1.6 (0.2) & 0.14 & 13.8 (3.0) & 2.0 \\
HD190073 & 9230 & 2.9 & -8.7 & -4.38 & 31. & 14. & 13.0 (4.7) & 0.30 & -- & 0.8 \\
HD244604 & 8700 & 2.8 & -7.2 & -4.31 & 55. & 49. & 2.7 (0.4) & <0.11 & -- & 1.4 \\

\noalign{\smallskip}
\textbf{\textit{GII disks}} & 8800 & 2.3 & -7.2 & -4.4  & 45  & 45  & 3  & 0.12  & 16  & 1.2 \\
 & (600) &  (0.5) &  (0.4) &  (0.2) &  (17) &  (14) &  (2) &  (0.04) &  (2) & (0.4) \\
\hline
\noalign{\smallskip}

HD31293 & 9800 & 2.5 & -7.7 & -4.87 & 24. & 14. & 7.6 (5.5) & 0.10 & 27.1 (2.9) & 4.5 \\
HD36112 & 8200 & 2.8 & -6.1 & -4.45 & 49. & 25. & 9.1 (5.7) & 0.02 & 27.5 (2.9) & 4.1 \\
HD100453 & 7250 & 1.5 & -8.0 & -4.57 & 49. & -- & -- & -- & 21.7 (2.7) & 5.2 \\
HD135344B & 6375 & 1.5 & -7.4 & -4.56 & 18. & 14. & 2.4 (1.1) & 0.05 & 27.2 (3.1) & 10.9 \\
HD142527 & 6500 & 1.6 & -7.5 & -4.59 & 33. & 28. & 2.1 (0.3) & 0.04 & 34.2 (3.3) & 5.0 \\

\noalign{\smallskip}
\textbf{\textit{high-NIR GI}} & 7300  & 2.0  & -7.5  & -4.6  & 33  & 20  & 5  & 0.05  & 27  & 5  \\
 &  (1300) &  (0.2) &  (0.3) &  (0.03) &  (21) &  (8) &  (4) &  (0.02) &  (0.4) &  (1) \\
\hline
\noalign{\smallskip}

HD34282 & 9500 & 1.9 & -7.7 & -5.30 & 66. & -- & -- & -- & 9.2 (1.0) & 10.0 \\
HD97048 & 10500 & 2.2 & -6.8 & -5.26 & 56. & 18. & 17.5 (2.3) & 0.30 & 9.8 (1.2) & 5.9 \\
HD100546 & 10400 & 2.3 & -7.0 & -5.67 & 46. & 17. & 14.6 (1.6) & 0.26 & 5.4 (0.9) & 3.5 \\
HD139614 & 7600 & 1.7 & -7.6 & -5.03 & 32. & 15. & 7.5 (3.4) & 0.25 & 7.8 (1.0) & 4.2 \\
HD141569 & 9800 & 2.4 & -7.7 & -5.21 & 53. & 16. & 20.9 (5.3) & 0.56 & 0.08 (0.01) & 6.8 \\
HD169142 & 7500 & 1.7 & -7.4 & -5.09 & 23. & 7. & 18.8 (15.0) & 0.23 & 10.5 (1.0) & 7.8 \\
HD179218 & 9640 & 3.1 & -6.7 & -4.99 & 49. & 20. & 15.7 (5.7) & 0.30 & 5.4 (0.8) & 2.4 \\
HD250550 & 11000 & 3.4 & -5.6 & -5.36 & 52. & 15. & 32.4 (12.8) & 0.16 & -- & 2.5 \\

\noalign{\smallskip}
\textbf{\textit{low-NIR GI}} & 9700  & 2.3  & -7.2  & -5.2  & 51  & 16  & 18 & 0.27  & 8  & 5  \\
 &  (1100) & (0.7) &  (0.7) &  (0.2) &  (7) &  (2) &  (4) &  (0.09) &  (4) &  (3) \\

            \hline
            \hline
         \end{tabular}
     $$ 

   \end{table*}

\section{ISHELL CO spectra} \label{sec:ishellspec}
\begin{figure}[ht]
\centering
\includegraphics[width=0.45\textwidth]{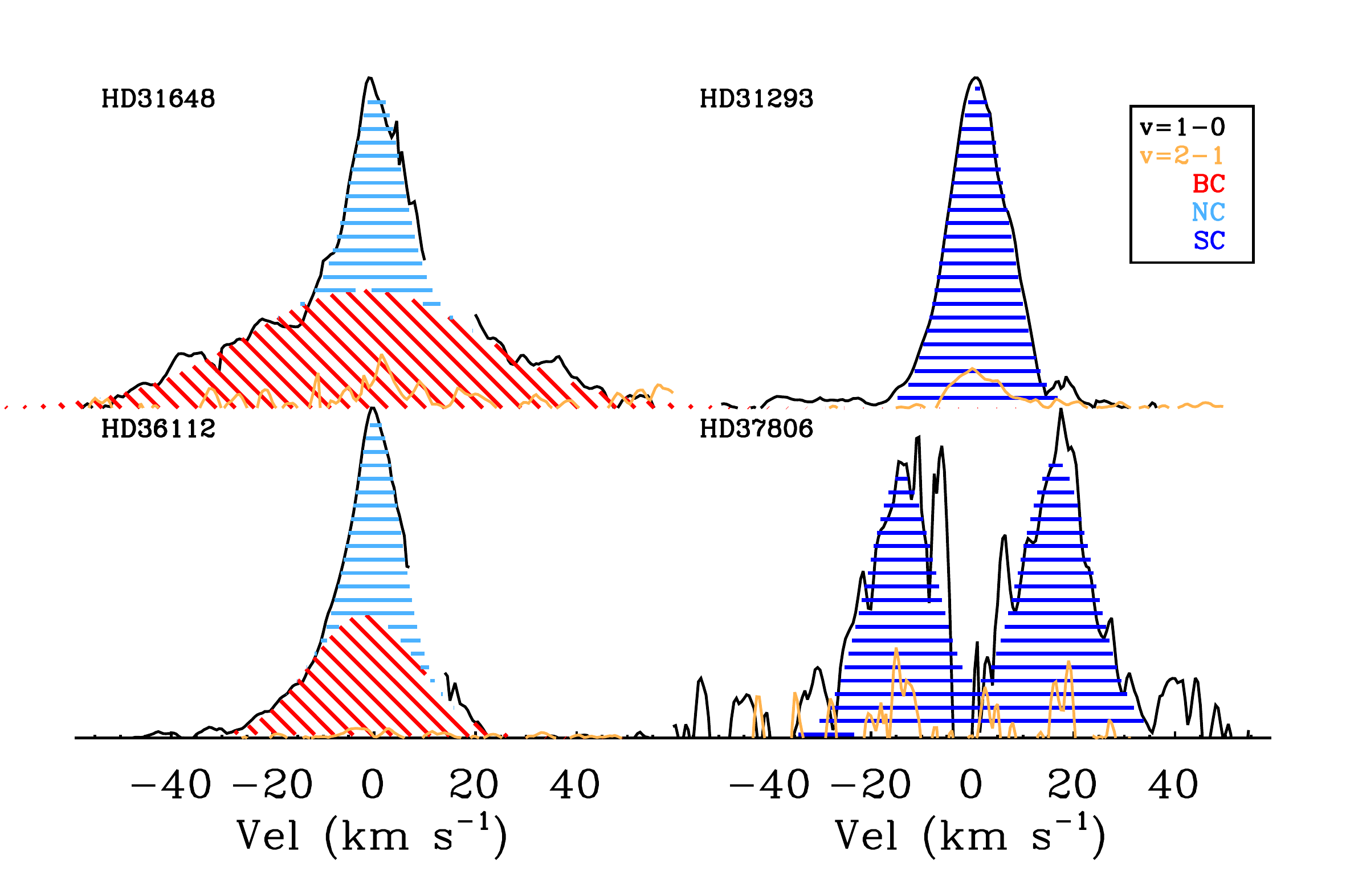} 
\caption{Stacked CO lines from the new ISHELL spectra, marking velocity components as in \citet{bp15}. HD31648 and HD36112 are the first Herbig disks found with two velocity components, which are typical of TTauri disks \citep{bp15}.}
\label{fig: ishell_lines}
\end{figure}

Four disks have been newly observed using IRTF-ISHELL \citep{ishell} in October 2016 (PI: Banzatti), providing a spectral resolution R $\sim75,000$, similar to VLT-CRIRES, which has been used for the other CO spectra. These disks are HD31293 (AB Aur), HD31648 (MWC 480), HD36112 (MWC 758), and HD37806. CO emission is detected in HD37806 for the first time. These spectra were reduced using a set of algorithms developed to reduce data from Keck-NIRSPEC, as described in \citet{britt07}, and adapted to ISHELL spectra. For this work, we have stacked the observed line profiles and measured FWHM$_{\rm co}$ as described in \citet{bp15}. The stacked lines are shown in Figure \ref{fig: ishell_lines} and can be directly compared to the rest of the CO line profiles published in \citet{bp15} and in \citet{vdplas15}.

\section{Modeling} \label{sec:model}
We adopted a simple disk model to explore how the NIR excess decreases as a function of the size of an inner dust-depleted disk region.
The model is based on physically motivated parameterizations and follows commonly used conventions. 
Inner disk radius and equilibrium temperature due to stellar irradiation are linked as

\begin{equation}
r = \left( \frac{{\rm L}_{\star}}{16\,\pi\,\sigma_{SB}\,T_{\rm d}^4}\frac{C_{\rm bw}}{\epsilon} \right)^{1/2},
\label{eq:rT}
\end{equation}

where the ratio of Planck mean opacities $\epsilon=\kappa_{\rm P}^{\prime}(T_{\rm dust})/\kappa_{\rm P}^{\prime}(T_{\star})$ is the dust cooling efficiency. We took $\epsilon=1$, which is appropriate where the dust is optically thick in the NIR and/or contains a significant amount of particles of size $\gtrsim1\,\mu$m. 
$C_{\rm bw} = 4$ is the back-warming factor for optically thick dust, and $4\pi/C_{\rm bw}$ is the spherically integrated solid angle available for cooling \citep{kama09}. 
The gas pressure scale-height is given by

\begin{equation}
h_{\rm gas} = \left( \frac{k_{\rm B}\,T\,r^{3}}{\mu\,{\rm m_{p}\,G\,M_{\star}}} \right)^{1/2}
.\end{equation}

The dust rim/wall has a surface area $A =  4\pi\,r\,H_{\rm h}\,h_{\rm gas}$, where $H_{\rm h}$ is a scalar factor specifying where the radial $\tau=1$ surface is reached in units of gas scale-height. We adopted $H_{\rm h} \sim 3$, in agreement with the height of the radial $\tau=1$ surface for stellar photons in the Monte Carlo radiative transfer and hydrostatic equilibrium models in the MCMax code \citep{min09, kama09}.
The dust rim/wall contributes 
\begin{equation}
L_{\rm w}=A\int_{c/5\,\mu m}^{c/1\,\mu m}B_{\nu}d\nu
\end{equation}
to the near-infrared excess. Beyond the rim, the disk surface is heated by the stellar flux modulated by the factor $\sin{(\beta)}$, with $\beta$ the angle between the ray and the local slope of a flared disk surface.
The local thermal balance and luminosity are analogous to the above, and the area per radial cell is $A_{\rm r}=2\pi r {\rm d}r$. The near-infrared luminosity from the disk surface is
\begin{equation}
L_{\rm s}=\int_{r_{\rm w}}^{r_{\rm out}}{ 2\pi r\int_{c/5\,\mu m}^{c/1\,\mu m}B_{\nu}\left[T(r)\right]d\nu dr}
\end{equation}
and the total NIR luminosity is $L_{\rm w}+L_{\rm s}$. 

\begin{figure*}[ht]
\centering
\includegraphics[width=0.7\textwidth]{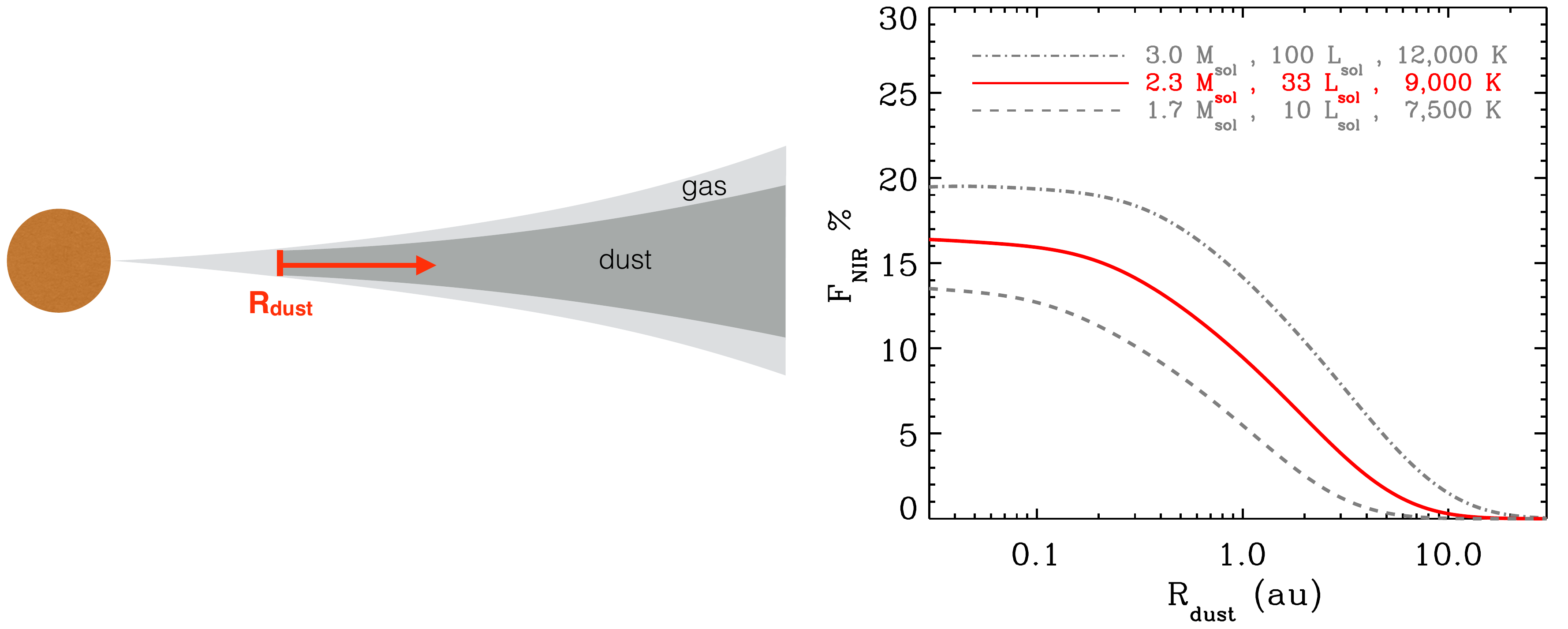} 
\caption{\textit{Left:} Illustrative cartoon of our simple modeling of inner holes with increasing size. \textit{Right:} The curves show the model dependence on stellar properties. In the reference model we take median stellar values for the sample (show in red, and reported in Fig.\ref{fig: correlations}). For comparison, we include two representative boundary cases.}
\label{fig: model}
\end{figure*}

We explored $\rm F_{NIR}$ as a function of dust radius R$_{\rm dust}$ by increasing the inner rim radius in the model, as illustrated in Figure \ref{fig: model}. $\rm F_{NIR}$ can be almost constant for small inner holes because of the balance between the increase in surface emitting area and the decrease in temperature. For larger inner holes, the temperature decreases faster than the surface area increases, and $\rm F_{NIR}$ drops.

\end{appendix}

\end{document}